\documentclass{article}
\usepackage{spconf,amsmath,graphicx,hyperref}

\usepackage{booktabs} 
\usepackage{subfigure} 

\usepackage{makecell}

\usepackage{tabularx}
\usepackage{multirow}
\usepackage{booktabs}
\usepackage{amssymb}
\usepackage{lipsum,cite}
\usepackage{amsmath}
\usepackage{bm}

\title{
A Distribution Matching Approach to Neural Piano Transcription \\
  with Optimal Transport
\vspace{-1mm}
}

\name{
\vspace{-1mm}
Weixing Wei\textsuperscript{1}
\quad
Raynaldi Lalang\textsuperscript{2}
\quad
Dichucheng Li\textsuperscript{3}
\quad
Kazuyoshi Yoshii\textsuperscript{2}
\vspace{-1mm}
\thanks{This work was supported 
by JST FOREST No. JPMJFR2270 
and JSPS KAKENHI Nos. 24H00742, 24H00748, 25K22841, and 25H01142.
Repo: \href{https://github.com/WX-Wei/AMT-optimal-transport}{https://github.com/WX-Wei/AMT-optimal-transport}.
}
}
\address{
\textsuperscript{1}Graduate School of Informatics, Kyoto University, Japan \\
\textsuperscript{2}Graduate School of Engineering, Kyoto University, Japan \\
\textsuperscript{3}Independent Researcher, Hong Kong, China
\vspace{-2mm}
}

\begin{document}

\begin{titlepage}
\thispagestyle{empty}

\vspace*{\fill}

© 20XX IEEE. Personal use of this material is permitted. Permission from IEEE must be obtained for all other uses, in any current or future media, including reprinting/republishing this material for advertising or promotional purposes, creating new collective works, for resale or redistribution to servers or lists, or reuse of any copyrighted component of this work in other works.

\vspace*{\fill}

\end{titlepage}

\fussy
\setlength{\abovedisplayskip}{4pt}
\setlength{\belowdisplayskip}{4pt}
\setlength{\abovedisplayshortskip}{3pt}
\setlength{\belowdisplayshortskip}{3pt}
\allowdisplaybreaks[4]
\maketitle

\begin{abstract}
\vspace{-1mm}
This paper describes a novel paradigm 
 that formalizes automatic piano transcription (APT)
 as an optimal transport (OT) problem,
 not as a frame-level multi-label binary classification problem.
Our method learns to minimize the cost of 
 transporting a predicted distribution of note events
 to the ground-truth distribution over time and frequency.
The OT loss can thus accommodate temporal misalignment,
 leading to perceptually relevant optimization.
We also propose a convolutional recurrent neural network (CRNN)
 with a harmonics-aware attention mechanism
 to capture the spectro-temporal dependencies inherent in music.
Our experiments using the MAESTRO dataset showed
 that our method attained a state-of-the-art performance
 in onset detection.
We confirmed the versatility of the OT loss 
 in application to existing models.



\end{abstract}


\vspace{-1mm}
\section{Introduction}
\vspace{-1.5mm}


%
In automatic piano transcription (APT)
 that aims to estimate a piano-roll representation (MIDI data)
 from a music recording~\cite{AMT-overview/benetos2018},
 deep learning models play a central role.
The standard paradigm is 
 to convert a music spectrogram over time frames and frequency bins
 into a piano-roll matrix 
 over time frames and semitone-level pitches
 through frame-level binary classification.
An APT model is trained to predict 
 the activations of the 128 MIDI pitches \cite{Onsets_Frames, high_resolution, 
 hppnet-Wei2022,hft-transformer-toyama2023}
 at each frame
 with the binary cross-entropy (BCE) loss.

The essential flaw of this frame-level classification is
 that the BCE loss treats each time-frequency bin 
 as an independent prediction.
A predicted note that may be misaligned 
 by a single time frame from the ground truth
 is severely penalized as a complete miss.
This temporal rigidity
 makes models overly sensitive to slight timing variations
 in performance or minor inaccuracies in dataset annotations.

 \begin{figure}[tb]
\centering
    \hspace{-3mm} 
    \includegraphics[width=7cm]{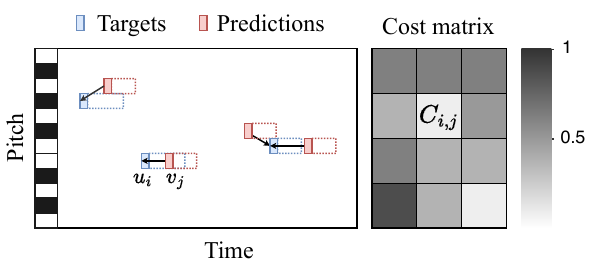}
    \vspace{-5mm}
\caption{Illustration of the optimal transport loss for piano transcription. 
(Left) A piano roll view of targets $u$ and predictions $v$.
(Right) The corresponding cost matrix $C_{i,j}$
encodes the pairwise distance between the target $u_i$ and prediction $v_j$.
}
 \label{fig:optimization-target-BCE-OT}
 \vspace{-5.5mm}
\end{figure}

To solve this fundamental problem,
 we formalize APT as distribution matching task
 with optimal transport (OT)
 \cite{villani2008optimal-transport-old-and-new}
 (Fig.~\ref{fig:optimization-target-BCE-OT}).
We treat the predicted notes with posterior probabilities
 as a mass distribution 
 over the time-frequency domain
 and the ground-truth notes
 as a set of target point masses.
The goal of model training 
 is to find the most efficient plan
 to \textit{transport} the predicted masses 
 to the target masses with the minimum cost.
This encourages the model 
 to place predictive masses
 near the target note events,
 naturally tolerating small timing variations. 
The transport cost is zero if and only if
 the predictions perfectly align with the ground truth.

In this paper,
 we explore the effect of the OT-based training in APT.
First, we propose a spatial-frequency-temporal 
 convolutional recurrent neural network (SFT-CRNN).
To make onset and offset predictions
 based on multifaceted dependency modeling over time and frequency,
 it uses convolutions over time,
 harmonics-aware self-attentions over frequency,
 and long short-term memory networks (LSTMs) and deconvolution 
 over time.
We then apply the OT-based training
 for our SFT-CRNN and existing models.
 
Our main contribution lies in
 the formalization of audio-to-MIDI transcription
 from the optimal transport perspective.
The OT-based training is model-agnostic 
 and can thus be easily used 
 by only replacing the BCE loss with the OT loss.
We show that our SFT-CRNN with the OT-based training
 achieved state-of-the-art results on standard AMT benchmarks,
 particularly for note onset detection.




\vspace{-2mm}
\section{Related Work}
\vspace{-2mm}


Early APT models \cite{SigtiaBD16-an-end-to-end-piano} 
 perform frame-wise classification with CNNs 
 to judge the presence of notes within each frame.
Onsets\&Frames model \cite{Onsets_Frames}
 predicted note onsets 
 and frame-wise activations separately 
 and then combining them afterwards.
A regression-based approach was proposed 
 for predicting high-resolution onset and offset times
 \cite{high_resolution}.
More recently, sequence-to-sequence models 
 with Transformers \cite{attention-is-all-you-need/nips/VaswaniSPUJGKP17} have been explored
 \cite{exploring-transformer/icassp/OuGBHW22, transformer-seq-to-seq-amt, MT3/iclr/GardnerSMHE22, Wei24-streaming-piano, icassp/LiZK25-piano-transcription-language-model}. 
Despite the theoretical advantage 
 in long-term dependency modeling inherent in music, 
 they may still underperform frame-level approaches 
 \cite{hppnet-Wei2022, hft-transformer-toyama2023, harmonic-transformer/icassp/wu, WangLBJ24-HarmonicawareFrequencyTime}.

Optimal transport has emerged 
 as a powerful tool for distribution matching
 \cite{arjovsky2017wasserstein-GAN, 
 cuturi2013sinkhorn-dist-lightspeed-optimal-transport}
 and has been applied to
 computer vision (e.g., 
 Wasserstein GANs\cite{arjovsky2017wasserstein-GAN}, 
 rectified-flow\cite{liu2022-rectified-flow})
  and natural language processing
  \cite{alvarez2018gromov-wasserstein-align-word-emb-spaces, xu2021vocabulary-learning-optimal-transport-machine-translate}.
%
In the MIR domain, 
 OT has shown promise for tasks requiring structural comparison. 
Wiering et al. 
 \cite{wiering2004transportation-distance-music-notation-retrieval} proposed transportation distance 
 for content-based music notation retrieval.
Flamary et al. \cite{nips/Flamary-AMT-Opt-Transport}
 proposed optimal spectral transportation 
 for music transcription.
Their approach framed transcription 
 as a spectral unmixing problem
 and operates directly on the spectrogram,
 aiming to decompose the spectral energy.
Instead of using OT for signal decomposition,
 we leverage it as a loss function 
 to evaluate the output of a neural network.

\vspace{-2mm}
\section{Proposed Method}
\vspace{-2mm}

This section introduces the problem formulation of APT 
 with OT and the proposed model architecture. 

\vspace{-3mm}
\subsection{Problem formulation}
\vspace{-1mm}

Let $\bm{X} \in \mathbb{R}^{T \times F}$ 
 be the time-frequency representation of a targe music recording,
 where $T$ is the number of frames 
 and $F$ is the number of frequency bins.
The ground-truth is a set of $N$ notes $G = {(s_i, e_i, p_i)}_{i=1}^N$,
 where $s_i, e_i, p_i$ are the onset time, offset time, 
 and pitch of the $i$-th note.


APT was conventionally formalized
 as frame-level binary classification.
The ground-truth $G$ is converted 
 to a piano-roll target $\bm{Y} \in \{0, 1\}^{T \times F'}$,
 where $F'$ is the number of pitches.
The model $\mathcal{M}$ predicts $\hat{Y} = \mathcal{M}(\bm{X})$,
 and the loss is the sum of per-frame binary cross-entropies:
$$\mathcal{L}_{B} =  - \sum_{t=1}^T \sum_{f=1}^{F'} [y_{t,f} \log(\hat{y}_{t,f}) + (1-y_{t,f}) \log(1-\hat{y}_{t,f})],$$
where $y_{t,f}$ and $\hat{y}_{t,f}$ represent
 the ground-truth and prediction 
 at frame $t$ and pitch $f$, respectively.



In our work, APT is formalized as distribution matching.
We define the ground-truth onset and offset events 
 as discrete distributions 
 (sum of two-dimensional Dirac delta functions)
 on the time-frequency grid as follows:
$$\bm{\mu}_{\text{on}} = \sum_{i=1}^N w \cdot \delta_{(s_i, p_i)}, \quad \bm{\mu}_{\text{off}} = \sum_{i=1}^N w \cdot \delta_{(e_i, p_i)},$$
where $w$ is a constant mass 
 assigned to each note (e.g., $w=1$)
 and $\delta$ is the Dirac delta function 
 defined as
$$\delta_{(a,b)}(x,y) = 0, (x-a)^2 + (y-b)^2 \neq 0,$$
$$
\iint_{\mathbb{R}^2}\delta_{(a,b)}(x,y) \, dx \, dy=1.
$$


Our model produces two non-negative outputs,
 the predicted onset mass distribution 
 $\bm{M_{\text{on}}} \in \mathbb{R}^{T \times F'}$
 and offset mass distribution 
 $\bm{M_{\text{off}}} \in \mathbb{R}^{T \times F'}$,
 where $T$ is the output temporal dimension.
The OT distance between 
 the predicted onset mass $\bm{M_{\text{on}}}$ 
 and the target $\bm{\mu_{\text{on}}}$ 
 is defined as
$$d_C(\bm{M_{\text{on}}}, \bm{\mu_{\text{on}}}) = \min_{\gamma \in \Pi(\bm{M_{\text{on}}}, \bm{\mu_{\text{on}}})} \int_{\mathbb{R}^2 \times \mathbb{R}^2} C(u, v) \, d\gamma(u, v),$$
where $u=(t_1, f_1)$ and $v=(t_2, f_2)$ 
 are points on the time-frequency grid, 
 $C(u, v)$ is the cost function,
 and $\Pi(\bm{M_{\text{on}}}, \bm{\mu_{\text{on}}})$ 
 is the set of all transport plans 
 (joint distributions) $\gamma$
 whose marginals are 
 $\bm{M_{\text{on}}}$ and $\bm{\mu_{\text{on}}}$.

In practice,
 we work with the discrete formulation of OT. 
Let $\bm{M}$ and $\bm{\mu}$ be vectorized representations
 of the predicted and target mass distributions
 over a grid of $D = T \times F'$ points.
The OT distance is the solution to the linear program:
$$d_C(\bm{M}, \bm{\mu}) = \min_{\gamma \in \Pi(\bm{M}, \bm{\mu})} \sum_{i=1}^{D} \sum_{j=1}^{D} \gamma_{i,j} C_{i,j} ,$$
where $C_{i,j}$ is the cost 
 of moving a unit of mass from grid point $u_i$ 
 to grid point $u_j$, 
 and $\gamma_{i,j}$ is the amount of mass 
 moved in the OT plan $\gamma$.
For balanced OT (BOT), the transport plans are 
subject to the constraints:
$$\Pi(\bm{M}, \bm{\mu}) = \{ \gamma \in \mathbb{R}_{+}^{D \times D} | \sum_{j=1}^{D} \gamma_{i,j} = \bm{M_i}, \sum_{i=1}^{D} \gamma_{i,j} = \bm{\mu}_j \}.
$$

\vspace{-5mm}
\subsection{Training strategy}
\vspace{-1mm}

We define a custom cost function $C(u_i, v_j)$ 
 that incorporates the practical realities 
 of music transcription.
Let $u_i = (t_i, f_i)$ and $v_j = (t_j, f_j)$ 
 be source and target grid points, respectively.
Our cost function follows two principles.
First, a temporal cost cap 
 limits the transport cost 
 for same-pitch events to a maximum value.
The cost should grow with the temporal distance 
 between the prediction and the target,
 but it should be capped 
 to prevent extreme gradient values 
 from single, distant mismatches.
We cap the cost at a maximum number, 
 $\tau_{0}$ (e.g., $\tau_0=5$).
Second, a predicted note should 
 never be matched to a ground-truth note 
 of a different pitch.
We enforce this with a large penalty 
 $\tau_{1}$ ($\tau_{1} \gg \tau_{0}$) for any transport 
 between different frequency bins.
This leads to the following cost function:
\begin{equation}
C'(u_i, v_j) \\  =
    \begin{cases}
    \min( |t_i - t_j|, \tau_{0} ) & \text{if } f_i = f_j, \\
    \tau_{1} & \text{if } f_i \neq f_j.
    \end{cases}
\end{equation}




 
For differentiability and computational efficiency, we further constrain the transport plan.
Each source point $u$ can transport mass to at most one target point $v$. This leads to a unique OT plan and the corresponding OT distance:
\begin{equation}
\gamma'_{i, j} \\  =
    \begin{cases}
    \bm{M}_i & \mbox{if} \ \  
    C'_{i,j} = \min_{j=1, \dots, \bm{D}} \{C'_{i,j}\}, \\
    0 & \text{otherwise},
    \end{cases}
\end{equation}
\\[-7mm]
\begin{equation}
    d_C'(\bm{M}, \bm{\mu}) 
    = \sum_{i=1}^{D} \sum_{j=1}^{D} 
    \gamma'_{i,j} C'_{i,j}.
\end{equation}

\begin{figure*}[tb]
    \begin{center}
    \includegraphics[width=160mm]{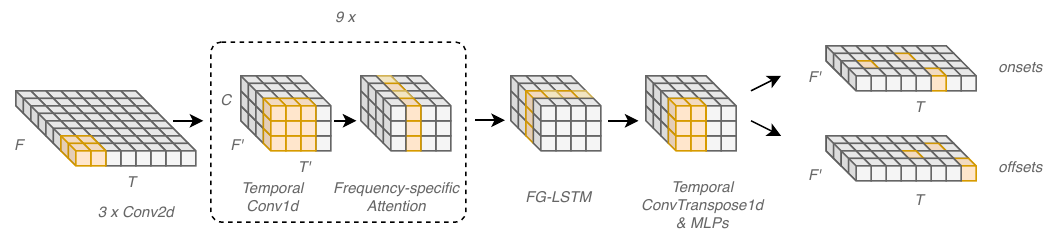}
    \end{center}
    \vspace{-7mm}
    \caption{The architecture of the proposed SFT-CRNN model. }
    \label{fig:SFT-CRNN}
    \vspace{-3mm}
\end{figure*}

In practice,
 we adopt unbalanced OT (UOT) \cite{chapel2021unbalanced-optimal-transport, sejourne2023unbalanced-optimal-transport-theory-to-numerics} rather than the balanced one
 for greater flexibility.
Balanced OT assumes that
 the source and target distributions carry equal total mass. 
Considering that the mass distribution of onsets and offsets in the piano-roll varies considerably with local note density, 
 we do not enforce this constraint.
Instead, we add an auxiliary mass penalty term that encourages the model output to
 match the target mass while not strictly enforcing equality:
\\[-4mm]
\begin{equation}
    \gamma'_{\text{max}}(j) =  \max_{i=1, \dots, D} \{\gamma'_{i,j}\} ,
\end{equation}
%
%
\\[-7mm]
\begin{equation}
\mathcal{L}_{\text{mass}} =  \sum_{j=1}^{D}  \left( \bm{\mu}_{j} - \gamma'_{\text{max}}(j) \right)^2 .   
\label{eq:loss-mass}
\end{equation}

The final OT loss is the sum of the transport cost and the mass penalty with a weighted factor (e.g., $\lambda=1$):
\\[-4mm]
\begin{equation}
    \mathcal{L}_{\text{OT}}(\bm{M}, \bm{\mu}) = d_C'(\bm{M}, \bm{\mu}) + \lambda\mathcal{L}_{\text{mass}} .
\end{equation}

The total loss for our model is given by
\\[-4mm]
\begin{equation}
    \mathcal{L} = \mathcal{L}_{\text{OT}}(\bm{M}_{\text{on}}, \bm{\mu}_{\text{on}}) + \mathcal{L}_{\text{OT}}(\bm{M}_{\text{off}}, \bm{\mu}_{\text{off}}) . 
\end{equation}

This loss intuitively penalizes 
 the model based on how far its predicted mass 
 is from the ground-truth locations,
 providing a smooth and perceptually 
 meaningful learning signal.




\vspace{-3mm}
\subsection{Model architecture}
\vspace{-1mm}

Our model, depicted in Fig. \ref{fig:SFT-CRNN},
 is a CRNN designed to
 effectively learn spectro-temporal features for the APT task.

\textbf{Convolutional blocks}: 
The constant Q-rransform (CQT)\cite{CQT} spectrogram 
 first passes through a stack
 of three 2D convolutional layers. 
Each layer uses a $7\times7$ kernel 
 and strides of (1,2), (1,2), and (2,1),
 progressively downsampling 
 the temporal and frequency dimensions 
 by factors of 2 and 4.
The channel count increases 
 from 1 to 64, 128, and finally 256.

\textbf{Harmonics-aware attention block}: 
It consists of nine stacked layers,
 each containing a temporal 1D CNN module 
  with a $7\times1$ kernel
  followed by a frequency-specific self-attention and
  multi-layer perceptron (MLP) modules. 
Each module has an instance normalization 
 and a residual connection.
The self-attention module is constrained
 to learn relationships 
 between harmonically related frequencies
 \cite{WangLBJ24-HarmonicawareFrequencyTime,
 harmonic-transformer/icassp/wu}.
An attention mask bias matrix $\bm{B}$ is pre-computed, 
 where $\bm{B}_{ij} = 0$ if frequency bin $i$ and bin $j$ 
 are harmonically related 
 (i.e., their fundamental frequencies are 
 close to an integer multiple of each other) 
 and $\bm{B}_{ij} = -\infty$ otherwise.
We alternate the harmonics-aware and full attentions
 to extract both 
 the harmonic and non-constrained dependencies.

\textbf{Temporal recurrent layers}: 
A frequency-grouped LSTM (FG-LSTM)
 is used for temporal dependency modeling
 \cite{hppnet-Wei2022}. 
Instead of using an LSTM 
 for the flattened feature map,
 we apply an independent LSTM at each frequency bin.
This prevents the smearing of features 
 across unrelated frequencies
 and allows the model to track the temporal pitch activities.

\textbf{Output heads}: 
The output of the FG-LSTM is upsampled 
 with temporal deconvolution by a factor of 2.
Two separate MLP branches 
 with the sigmoid activation functions 
 predict the final onset mass distribution $M_{\text{on}}$
 and offset one $M_{\text{off}}$.

\vspace{-3mm}
\section{Evaluation}
\vspace{-2mm}

This section reports comparative evaluation
 conducted for validating the effectiveness of the OT loss
 in APT.
 
\vspace{-3mm}
\subsection{Experimental Conditions}
\vspace{-1mm}

The MAESTRO\cite{maestro} dataset was used for evaluation.
It contains over 200 hours of piano recordings
 with aligned MIDI data.
We used the official train/validation/test split.
Each recording was resampled at 48 kHz, 
 clipped to a duration of 10 seconds, 
 and converted to the CQT spectrogram 
 with 352 frequency bins, 48 bins per octave, 
 a hop length of 1200, 
 and a minimum frequency of 27.5 Hz.
We used the Adam \cite{Adam} optimizer 
 with a learning rate of $10^{-4}$ for training.


We computed the precision and recall rates
 and F1-score for evaluating the performance
 in terms of onset estimation only
 or both onsets and offset estimations
 with the mir\_eval library \cite{mir_eval}.
The onset times predicted within a 50 ms tolerance window 
 were considered correct.
The offset times predicted within a 50 ms tolerance window 
 or 20\% of the note duration
 were considered correct.
 

\vspace{-3mm}
\subsection{Comparative Study}
\vspace{-1mm}

We compared our model with the well-established baselines.
As shown in Table \ref{tab:maestro},
 our method established a new state-of-the-art in onset F1-score (98.36\%), surpassing all evaluated baselines.
This result clearly demonstrated the effectiveness of our approach.
For the overall note transcription task (evaluating onsets and offsets), our model achieved a highly competitive F1-score of 90.78\%.
This performance is excellent,
 though it does not exceed the score of the best-performing system on the specific metric.
We hypothesize that this is because note offsets are influenced by two factors:
 the physical release of the piano key and the sustain pedal.
Our current model did not explicitly detect pedal events yet.
We believe that incorporating a pedal processing module is a promising direction for further improving offset accuracy.

\begin{table}[t]
 \centering
 \caption{Comparison of the proposed and compared methods.}
 \label{tab:maestro} 
 \vspace{-3mm}
 \ninept
 \begin{center}
 \hspace{-0.11in} 
 \resizebox{.83\width}{!}{
 \renewcommand{\arraystretch}{1.3}
 \setlength{\tabcolsep}{3.5pt}
 \begin{tabular}{l c c c c c c c c c c }
 \toprule
 \multirow{2}{*}{Model} & \multirow{2}{*}{Params} & \multicolumn{3}{c}{Onset} &
 \multicolumn{3}{c}{Onset \& Offset} \\
 \cmidrule(r){3-5} \cmidrule(r){6-8} 
 
 &  & P (\%) & R (\%) & F1(\%) 
 & P (\%) & R (\%) & F1(\%) \\
 
\midrule

Onsets \& Frames \cite{maestro} & 26M & 98.27 & 92.61 & 95.32 &
82.95 & 78.24 & 80.50  \\


 
%
HPPNet-sp \cite{hppnet-Wei2022} & 1.2M & 98.45 & 95.95 & 97.18 & 84.88 & 82.76 & 83.80  \\

hFT-Transformer \cite{hft-transformer-toyama2023} & 5.5M &  99.64 & 95.44 & 97.44 & 92.52 & 88.69 & 90.53 \\

Transkun \cite{yan2024scoring} & 12.9M &  99.53 & 97.16 & \underline{98.32} & 94.61 & 92.39 & \textbf{93.48} \\



\hline
SFT-CRNN & 15M &  99.16 & 97.46 & \textbf{98.36} & 91.56 & 90.02 & \underline{90.78}  \\

 
 \bottomrule
 \end{tabular} 
 }

\end{center}

\vspace{-7mm}
\end{table}


\vspace{-3mm}
\subsection{Ablation Study}
\vspace{-1mm}


To evaluate the effectiveness of the proposed OT loss 
 in different model architectures,
 we performed additional experiments on the proposed model, the Onsets \& Frames model\cite{maestro} and the HPPNet-base model \cite{hppnet-Wei2022}, 
 using the OT loss and the BCE loss for training separately.
The Onsets \& Frames model was trained 
 with the same configuration as in the original paper,
 while for HPPNet-base, the maximum number of channels 
 was increased from 128 to 256 to ensure the sufficient model capacity.
As for the BCE loss, the onset/offset lengths 
 in the target piano-roll were set to 2 frames.

\begin{table}[t]
    \centering
    \caption{F1 scores of onset estimation.}
    \label{tab:ablation-loss-function} 
    \vspace{1mm}
    \ninept
    \resizebox{.85\width}{!}{
        \begin{tabular}{llcc}
        \toprule
        Model & Loss   & Onset & Onset \& Offset   \\
        
        \midrule
        \multirow{2}{*}{Onsets \& Frames \cite{maestro}} & BCE Loss  & \textbf{96.21} & 78.71  \\
        & OT Loss  & 96.15 & \textbf{79.33} \\
        
        \midrule
        \multirow{2}{*}{HPPNet-base \cite{hppnet-Wei2022}} & BCE Loss  & 97.03 & 85.71  \\
        & OT Loss  & \textbf{97.49} & \textbf{87.38} \\
        
        \midrule
        \multirow{2}{*}{SFT-CRNN} & BCE Loss  & 97.61 & 88.58  \\
        & OT Loss  & \textbf{98.36} & \textbf{90.78} \\
        \bottomrule
        \end{tabular}
        }
        \vspace{-5mm}
\end{table}

 

As shown in Table \ref{tab:ablation-loss-function},
 replacing the OT loss with the BCE loss 
 led to a clear performance drop in the SFT-CRNN model,
 where the onset F1 score decreased from 98.36\% to 97.61\%, 
 and the onset \& offset F1 score decreased from 90.78\% to 88.58\%. 
This result provided direct evidence that
 the OT loss is a key driver of our performance gains.
A similar trend was also observed 
 for the HPPNet-base model.
In contrast, the Onsets \& Frames model exhibited 
 little performance difference between the two loss functions.
This was attributed to the performance limitations of this model,
 which may have dominated the overall performance 
 and reduced the relative impact of the loss functions.

 \begin{figure}[tb]
   \includegraphics[width=8.6cm]{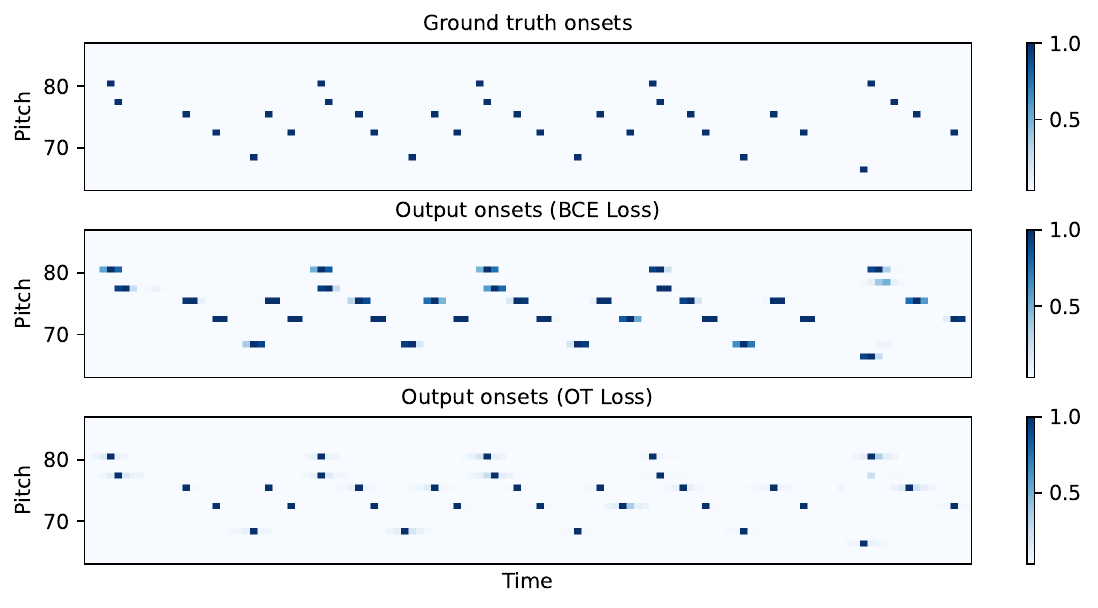}
   \vspace{-8mm}
    \caption{The note onsets predicted 
    by the proposed SFT-CRNN model 
    trained with BCE loss or OT loss.}
    \label{fig:onset_pianoroll_with_different_losses}
    \vspace{-2mm}
 \end{figure}

Beyond quantitative metrics, 
 the qualitative nature of the raw model output 
 offered significant insight 
 into the model behavior and practical utility.
Fig.~\ref{fig:onset_pianoroll_with_different_losses} 
 directly compares a typical onset piano-roll generated
 by the proposed SFT-CRNN model trained with the standard BCE loss 
 and one trained with the proposed OT loss.
We observed that the model trained with the BCE loss 
 tended to produce activations smeared across several consecutive frames 
 around the onset of a single note.
This well-known characteristic of frame-level classification models
 necessitates a post-processing step, 
 typically involving peak-picking within a time window, 
 to avoid detecting multiple onsets for a single true note.
In contrast, the output activations 
 from the model trained with the OT loss were remarkably sharper 
 and concentrated into single frames, 
 precisely aligning with the ground truth onsets. 
This inherent quality represents a major practical advantage 
 of the OT loss for predicting discrete events.

\begin{table}[tb]
    \centering
    \caption{Ablation study on the proposed SFT-CRNN model.}
    \label{tab:ablation-model}
    \vspace{1mm}
    \resizebox{.88\width}{!}{
        \begin{tabular}{lcc}
        \toprule
        Model    & Onset & Onset \& Offset   \\        
        \midrule
        {SFT-CRNN (proposed)}  & \textbf{98.36} & \textbf{90.78}  \\
        w/o LSTM  & 97.74 & 86.68 \\
        w/o harmonics-aware attention & 97.66 & 87.56 \\
        \bottomrule
        \end{tabular}
    }
    \vspace{-3mm}
\end{table}

To verify the effectiveness of each component within the proposed SFT-CRNN, 
 we conducted an ablation study 
 by systematically removing the LSTM module 
 and the harmonic-aware attention mechanism.
As shown in Table \ref{tab:ablation-model}, 
 removing the LSTM led to a 4.10 pts reduction in the onset/offset F1-score, 
 which strongly indicates its crucial role in temporal modeling. 
Similarly, the removal of the harmonic-aware attention mechanism 
 resulted in a 3.22 pts drop, 
 confirming its distinct effectiveness. 
These results demonstrate that both components are essential 
 for achieving the strong overall performance of the full proposed model.




\vspace{-2mm}
\section{Conclusion}
\vspace{-2mm}


This paper introduced a novel paradigm 
 that formalizes APT as an OT problem, 
 moving beyond the temporal rigidity limitations 
 of traditional frame-level binary classification problem. 
We proposed a new CRNN model
 to capture the spectro-temporal dependencies of music
 and trained it in a perceptually reasonable manner 
 to accommodate temporal misalignment.
The OT loss is model-agnostic, making it easily applicable 
 to existing models.

Future work includes exploring 
 a dedicated sustain pedal detection module 
 to enhance offset accuracy.
We also plan to apply this powerful OT framework 
 to other music information 
 retrieval tasks such as drum transcription, 
 chord recognition, and sound event detection.





\ninept
\bibliographystyle{IEEEbib}

\bibliography{strings,refs}

\end{document}